\documentclass[aps,prl,reprint,a4paper]{revtex4-1}
\usepackage{graphicx}
\usepackage{amsmath}
\newcommand{\tsub}[1]{$_\text{#1}$}
\newcommand{\tsuper}[1]{$^\text{#1}$}

\begin{document}
%\preprint{Draft }
\title{Critical behavior of sputter-deposited magnetoelectric antiferromagnetic Cr\tsub{2}O\tsub{3} films near N\'eel temperature} 
\author{Muftah Al-Mahdawi}
\email{mahdawi@ecei.tohoku.ac.jp}
\author{Yohei Shiokawa}
\author{Satya Prakash Pati}
\author{Shujun Ye}
\author{Tomohiro Nozaki}
\author{Masashi Sahashi}
\affiliation{Department of Electronic Engineering, Tohoku University, Sendai 980-8579, Japan}
\date{\today}
\begin{abstract}
Chromium(III) oxide is a classical collinear antiferromagnet with a linear magnetoelectric effect.
%However, most of studies on the phase-transition of thin-films were indirect inferences from exchange-bias on .
 We are presenting the measurements of the magnetoelectric susceptibility $\alpha$ of a sputter-deposited 500-nm film and a bulk single-crystal of Cr\tsub{2}O\tsub{3}. We investigated the magnetic phase-transition and the critical exponent $\beta$ of the sublattice magnetization near N\'eel temperature. For the films, an exponent of 0.49(1) was found below 293 K, and changed to 1.06(4) near the N\'eel temperature of 298 K. For the single-crystal, the exponent was constant at 0.324(4). We investigated the reversal probability of antiferromagnetic domains during magnetoelectric field cooling. For the sputtered films, reversal probability was zero above 298 K and stabilized only below 293 K. We attribute this behavior to formation of grains during film growth, which gives different intergrain and intragrain exchange-coupling energies. For the single-crystal, reversal probability was stabilized immediately at the N\'eel temperature of 307.6 K.
\end{abstract}
\pacs{}
\maketitle
\section{Introduction}
One of the most important characters of antiferromagnets (AFMs) is the lack of an observable magnetization. The atomic magnetic moments from each sublattice compensate each other due to an antiferromagnetic coupling mediated by the exchange interaction. Therefore, AFMs are stable to the applied magnetic fields, which is advantageous for data retention and protection, but problematic for reading and writing. One way is to utilize the exchange coupling with a ferromagnetic layer as a readout, and control AFM domains by another external stimulus, \emph{e.g.}~electric field in magnetoelectric media \cite{he_2010,heron_2014}. The linear magnetoelectric (ME) effect is the reciprocal effect of inducing a magnetization $M$ linearly proportional to an applied electric field $E$ and an electric polarization due to a magnetic field \cite{odell_1970-1}. The typical antiferromagnetic magnetoelectric material chromia (Cr\tsub{2}O\tsub{3}) gained a renewed interest as a strong candidate for electric-writing schemes \cite{borisov_2005,chen_2006-1,ashida_2014,toyoki_2015,he_2010,ashida_2015,toyoki_2015-1,shibata_2015_edited}. This is due to the potential of room-temperature operation \cite{he_2010}, and the progress in growing high-quality epitaxial films \cite{shiratsuchi_2011,pati_2015}.\\
Chromia is an AFM with a corundum-type crystal structure where spins are aligned parallel to the \emph{c}-axis. The alternating spin configuration of $+$ $-$ $+$ $-$ (Fig.~\ref{fig:exp}(a)) breaks space-inversion and time-reversal symmetries, and results in a linear ME effect as predicted by Dzyaloshinskii \cite{dzyaloshinskii_1960}. and confirmed experimentally by Astrov \cite{astrov_1960}. There are two 180$^\circ$-related possible domains $F^\pm$, which are stabilized depending on the relative orientation of magnetic and electric fields during field cooling (MEFC). The Cr spins in $F^+$ ($F^-$) domain point inward (outward) the shared smaller oxygen triangle between oxygen octahedra (shaded area in Fig.\ref{fig:exp}(a)) after a $+$MEFC ($-$MEFC) \cite{martin_1964,brown_1998}.\\
The second-order phase-transition of AFMs is same as ferromagnets, except that the staggered-moment vector (N\'eel vector) is the order-parameter instead of the macroscopic magnetization. The critical exponent $\beta$ of order-parameter of Cr\tsub{2}O\tsub{3} films was reported before \cite{borisov_2011,fallarino_2015}. Inferences from exchange-bias effect on a proximate ferromagnet \cite{borisov_2011}, and magnetization from uncompensated surface spins \cite{fallarino_2015} were used to determine $\beta$ at the surface of Cr\tsub{2}O\tsub{3} near N\'eel temperature $T_N$. It was argued that Cr\tsub{2}O\tsub{3} have the character of 2D surface magnetism near $T_N$ \cite{borisov_2011,fallarino_2015}, due to a cross-over from a 3D-bulk behavior to a 2D-surface one \cite{binder_1974,binder_1984}.\\
In this report, we measured the magnetoelectric susceptibility $\alpha$ during the transition to the paramagnetic (PM) phase near $T_N$, which is proportional to the volume-averaged sublattice magnetization ($\alpha \propto \langle M_\mathrm{sub}\rangle$) \cite{rado_1961}. For a single-layer film of Cr\tsub{2}O\tsub{3}, we measured $\beta$ near $T_N$ and found a cross-over of $\beta$ and magnitudes different from the bulk-substrate sample. Also, we investigated the relation between the temperature at which $\beta$ changed in magnitude, and the domain reversal during field cooling.\\
\section{Experiment}
\begin{figure}
	\caption{(a) Ions of Cr$^{3+}$ (blue spheres with spin direction) and O$^{2-}$ (red spheres) crystallize in a corundum-type crystal. $F^\pm$ domain state with positive (negative) magnetoelectric susceptibility, $\alpha = dM/dE$, forms after a positive (negative) magnetoelectric field cooling (MEFC). (b) A schematic of sample geometry and directions of electric and magnetic fields. The electric field is applied only at the cross-junction denoted by a cube, whereas the magnetic field is uniform. A photograph of the sample is shown in the inset. (c) Two lock-in amplifiers (LIA) are used. One is connected to the flux-locked-loop output to detect the electrically-induced magnetization, and the second to a trans-impedance amplifier that measures the leakage current. (d) The experimental procedure of the effect of MEFC stopping temperature $T^\star$. The electric and magnetic fields ($E_\mathrm{fr}$, $H_\mathrm{fr}$) are set to zero below $T^\star$. Then $\alpha$ at its peak $\alpha_\text{peak}$ is measured at the corresponding temperature $T_\text{peak}$.}
	\label{fig:exp}
	\includegraphics[width=1.0\textwidth]{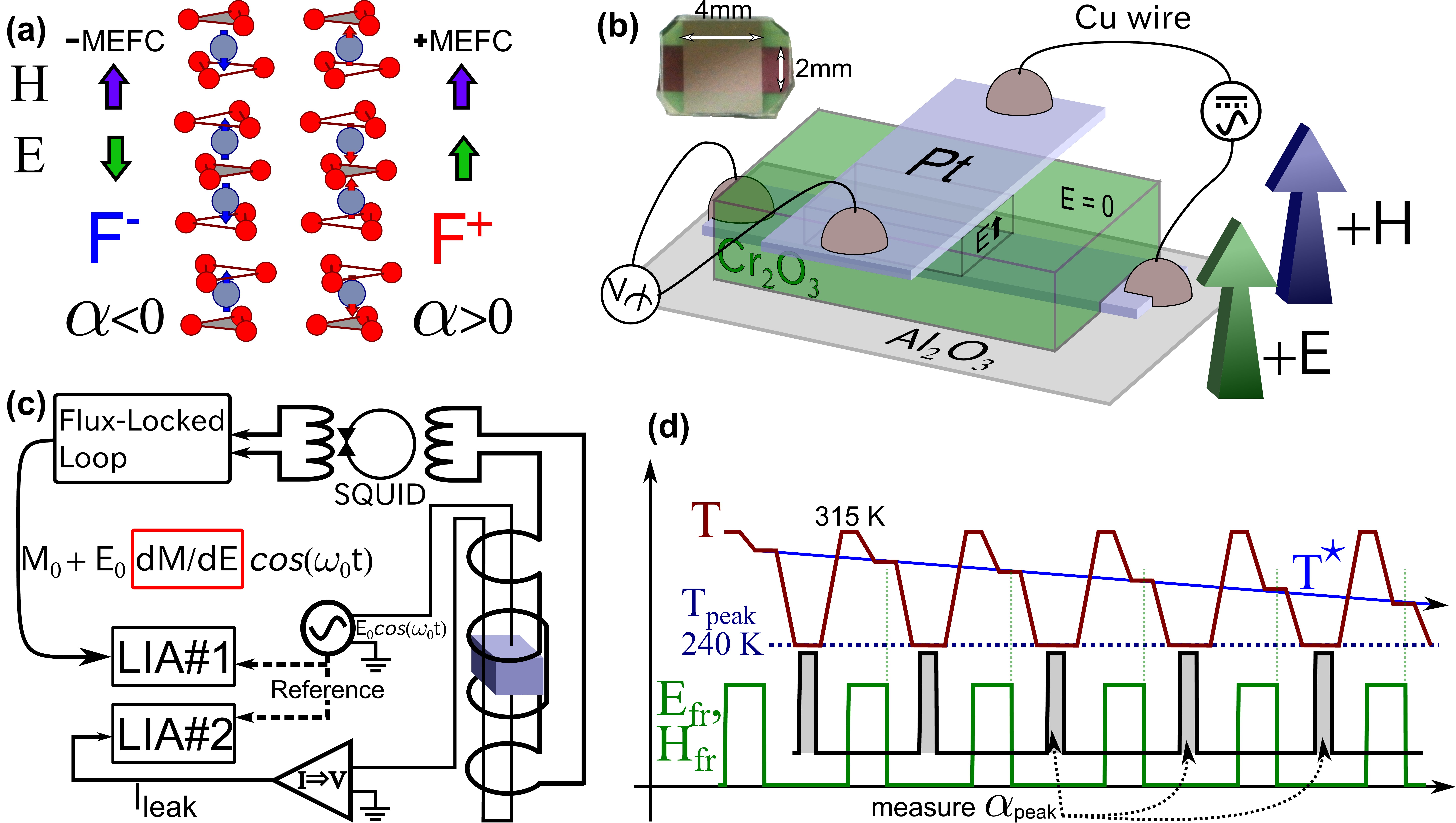}
\end{figure}

Two samples were used: a 4$\times$4$\times$0.5-mm\tsuper{3} commercial 3N Cr\tsub{2}O\tsub{3} substrate made by Verneuil process from Furuuchi Chemical, and a sputter-deposited Cr\tsub{2}O\tsub{3} film. The presented results are from a single sample for each, and they were qualitatively reproducible for other ones. The sputtered-film structure was as follows: \emph{c}-Al\tsub{2}O\tsub{3} (0001) substrate/Pt (25 nm, 773 K)/Cr\tsub{2}O\tsub{3} (500 nm, 773 K)/Pt (25 nm, 423 K), where temperature values are the substrate temperatures during deposition. The optimized growth conditions for previous reports were used \cite{ashida_2015,borisov_2016}. The growth characterization of Cr\tsub{2}O\tsub{3} over a Pt buffer was reported before \cite{pati_2015,borisov_2016,shimomura_2016}, and \emph{c}-axis-oriented epitaxial growth was confirmed. However, Cr\tsub{2}O\tsub{3} is strained and 30$^\circ$-rotated domains are embossed from the Pt buffer. The in-plane strain was quantified at 0.6\%, but the out-of-plane strain was unknown due to the overlap of Pt (111) and Cr\tsub{2}O\tsub{3} (0006) x-ray-diffraction peaks. The thicknesses were determined from calibrated deposition rates of 4.6 and 3.8 nm/min for Pt and Cr\tsub{2}O\tsub{3}, respectively.
%The error in thickness was usually on an order of $<$2\%, as determined from transmission electron micrographs of other samples.
 The choice of Cr\tsub{2}O\tsub{3} thickness was to minimize the leakage current during measurement. The bottom Pt electrode was prepared by sputtering through a slit in a stencil mask, then the mask was removed in air for subsequent deposition. Cr\tsub{2}O\tsub{3} was deposited from a 4N Cr-metal target by RF-magnetron reactive sputtering, with the oxygen flow controlled by a plasma emission monitor. An orthogonal top Pt electrode was made by photolithography and ion etching. The electric field was applied at the cross-junction and the magnetic field was applied uniformly on the whole substrate. The area of crossing was designed at 8 mm\tsuper{2} and was measured by an optical microscope. A schematic of the device geometry and field directions is shown in Fig.~\ref{fig:exp}(b).\\
The linear ME \emph{ac} susceptibility $\alpha = dM/dE$ was measured from the magnetization induced by an \emph{ac} electric field, using a commercial Quantum Design MPMS-XL SQUID magnetometer (Fig.~\ref{fig:exp}(c)) \cite{borisov_2007,borisov_2016}. The electric field was applied by an external \emph{ac} source connected through a sample holder equipped with electrical connections. The induced \emph{ac} magnetization was detected by a phase-sensitive detection of the flux-locked-loop output using a dedicated lock-in amplifier. However, a part of the detected magnetization comes from the leakage current flowing in the wire loop made by the sample and connections. The leakage current was simultaneously measured by a low-noise transimpedance amplifier for background correction. The readings of current were used to compensate for the background with a single proportionality constant. This compensation factor was chosen to zero the detected magnetization above $T_N$, and a single value could be used during the span of experiments. This scheme allowed keeping the sample at the center of gradiometer pick-up coil for weeks, without the need to move the holder for nulling at null-response points \cite{borisov_2007}, and no drift was observed even after two weeks of various experiments. The excitation frequency was determined by searching the spectrum for the least noise while keeping the capacitive current at a minimum, and was fixed at 9 Hz. Peak-to-peak noise was less than $5\times10^{-9}$ emu at a zero magnetic field. Sample centering was done at a high field by the substrate's diamagnetic response.\\
In the first experiment, the temperature dependence of $\alpha$ was measured after MEFC at a freezing magnetic field $H_\mathrm{fr}$ and a freezing \emph{dc} voltage $V_\mathrm{fr}$ from 315 K to 30 K. Measurement was done under a zero magnetic field, an \emph{ac} voltage of 8.5 V peak-to-peak, and a 0.5-K/min heating rate. In a second experiment, we explored the relation between the domain-reversal probability and the transition region of 293--298 K around $T_N$. The effect on reversal probability by the sign of MEFC and at which temperature $T^\star$ MEFC is stopped was investigated. Ideally, the average domain state should be measured directly after setting magnetic and electrical fields to zero at $T^\star$. However, the amplitude of $\alpha$ is very small near $T_N$. So after MEFC from 315 K to $T^\star$, magnetic and electric fields are set to zero, then the domain state is zero-field-cooled to the peak temperature of $\alpha$. At the peak temperature, $\alpha_\text{peak}$ was measured for 3 min. A schematic of the procedure is shown in Fig.~\ref{fig:exp}(d). Same procedures were followed for the Cr\tsub{2}O\tsub{3} single-crystal substrate, except MEFC was from 330 K and measurement of $\alpha_\text{peak}$ was at 270 K with a sense voltage of 200 V peak-to-peak. We minimized the temperature undershoot at $T^\star$ to $<$0.25 K during MEFC from a high temperature.\\
\section{Results and discussion}
\begin{figure}
	\caption{(a) Temperature dependence of $\alpha$ and leakage current in thin-film Cr\tsub{2}O\tsub{3} after positive and negative MEFC. Corrected and uncorrected measurements from magnetic field background of leakage current are marked by a solid triangle and an empty square, respectively. (b) The critical exponent $\beta$ changed near $T_N$ of the Cr\tsub{2}O\tsub{3} film, whereas it stayed constant for the bulk substrate. The fitting ranges are indicated by arrows next to each $\beta$ value, with the corresponding fittings (dashed lines) extrapolated.
	%(c) The cross-over of $\beta$ with temperature near $T_N$ from 0.486 to 1.061. The quality of fit is indicated by the adjusted coefficient of determination ($\bar{R}^2$). (d) The estimated $T_N$ also changed from $\approx$293 K to $\approx$298 K. In (c) and (d), vertical error-bars are the uncertainty in fitting, and horizontal bars are the fitting ranges.
	}
	\label{fig:a-T}
	\includegraphics[height=0.5\textheight]{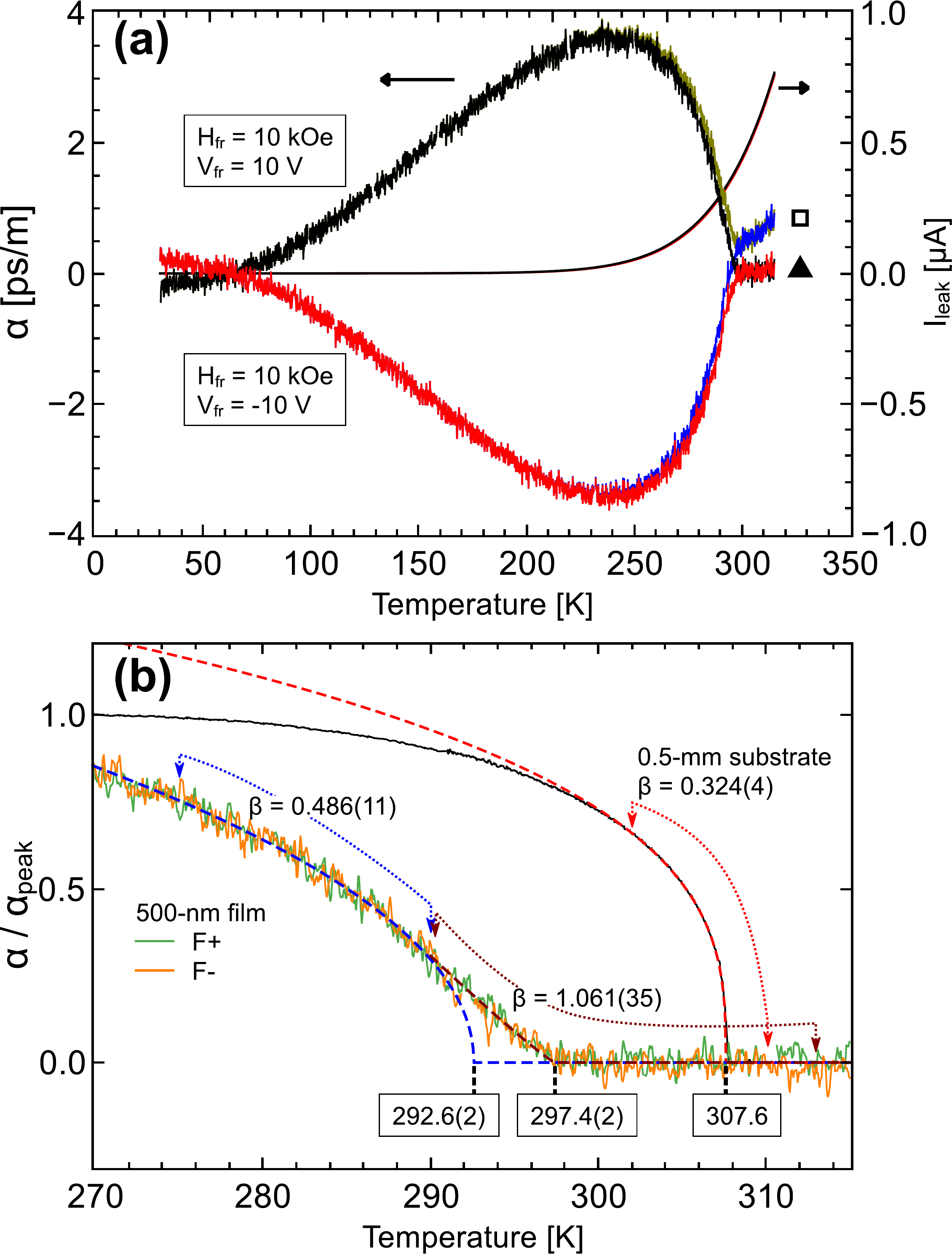}
\end{figure}

\begin{figure}
	\caption{The dependence of switched domain population on $T^\star$ in (a) the film sample, and (b) the bulk substrate. Error bars for $\langle F \rangle$ denote the standard deviation of averaged measurements, and for temperature denote undershoot during cooling towards $T^\star$. Solid lines are eye-guides.\\}
	\label{fig:T-star}
	\includegraphics[height=0.5\textheight]{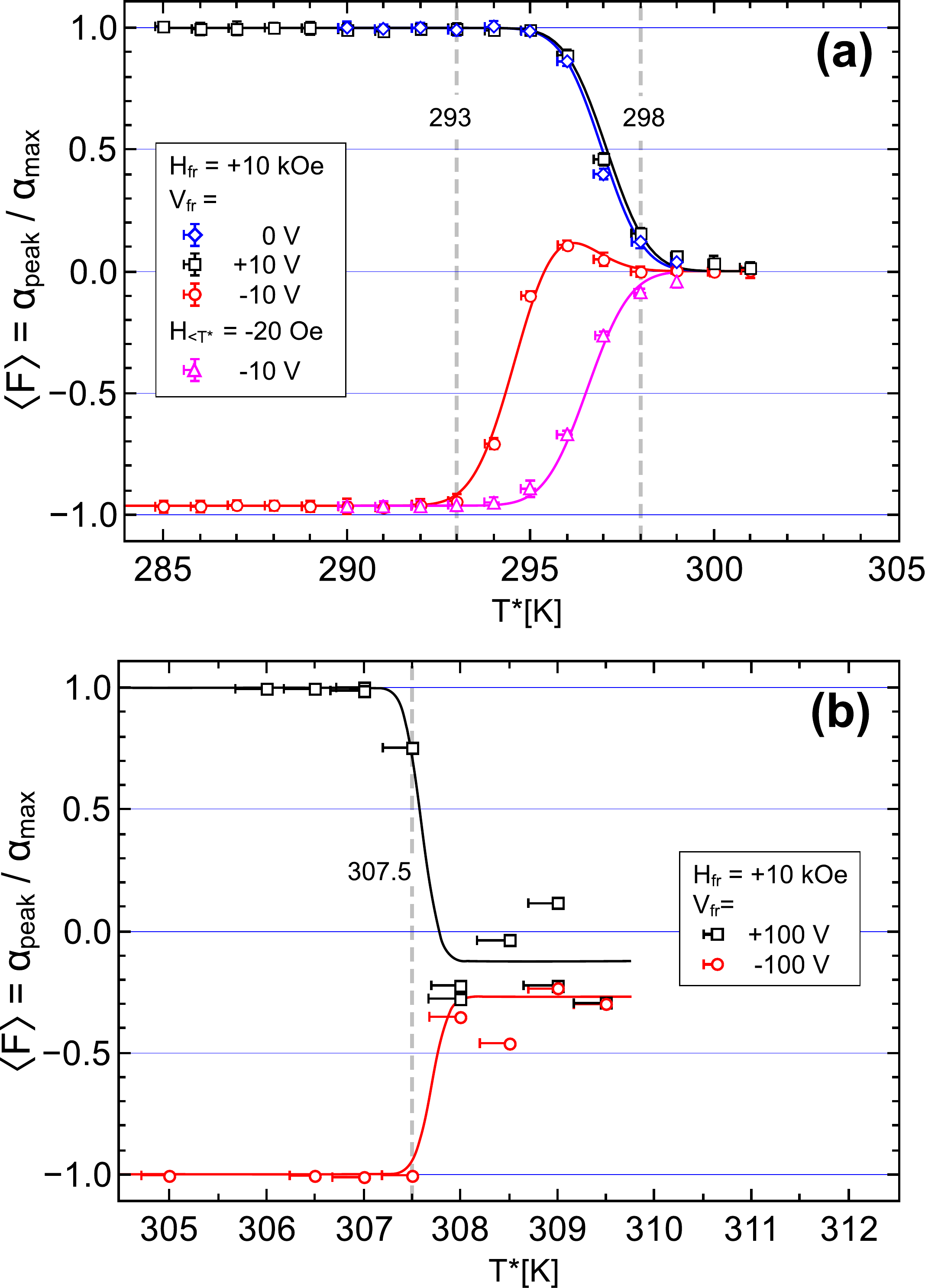}
\end{figure}

The temperature dependence of the leakage current (Fig.~\ref{fig:a-T}(a)) showed an increase with increasing temperature, indicating a semiconducting character. Still, the resistance was high (4 M$\Omega$) and a uniform electric field can be assumed along the thickness of the film.
%The electrodes and connections resistance was $\approx$ 50 $\Omega$, and 
 The voltage measured by the four-wire method (Fig.~\ref{fig:exp}(b)) was equal to the applied voltage up to 10 V. Thus, the electric field distribution is uniform across the area of the film.\\
The wide-range temperature dependence of magnetoelectric susceptibility $\alpha$ after $\pm$MEFC is shown in Fig.~\ref{fig:a-T}(a). The measurements without correction for leakage background and the ones with correction are denoted by a square and a triangle in Fig.~\ref{fig:a-T}(a), respectively. After correction, $\alpha$-T dependence after positive and negative MEFC became symmetric.
 The maximum peak value of $\alpha$ in the film sample $\alpha_\text{max}$ of 3.6 ps/m is same as the bulk value \cite{borisov_2007}, within errors in sample misalignment and centering. A similar result was recently reported \cite{borisov_2016}, and indicates that the magnetoelectric properties of the bulk crystals and sputtered films are similar.\\
The positive (negative) MEFC at $+$10 kOe and $+$10 V ($-$10 V) resulted in the saturation of $\alpha_\text{peak}$ at $+\alpha_\text{max}$ ($-\alpha_\text{max}$), indicating the formation of a single $F^+$ ($F^-$) AFM domain state. MEFC at $+$10 kOe and 0 V was identical to $+$10 V, showing that the sputter-deposited Cr\tsub{2}O\tsub{3} films respond to the applied magnetic field \cite{nozaki_2014,borisov_2016}.
% The lower noise and leakage correction attained in this report allowed for more accurate measurements around $T_N$.
 The temperature dependence of $\alpha$ was same for $F^+$ and $F^-$ states (Fig.~\ref{fig:a-T}(b)). The average of four temperature sweeps after MEFC at different $H_\mathrm{fr}$'s and $V_\mathrm{fr}$'s was used for the subsequent fitting. The sublattice magnetization's critical exponent $\beta$ of the film sample was determined by fitting the magnetoelectric susceptibility to a power law $A\cdot[(1-T/T_N)\cdot u(1 - T/T_N)]^\beta$ at different temperature ranges near $T_N$, where $u(x)$ is the Heaviside step function.
% The Levenberg-Marquardt algorithm was used for the least-squares fitting.
 The fitting could not be done using a single pair of $\beta$ and $T_N$ values. They changed from 0.486(11) and 292.6(2) K at a fitting range of 275--290 K, to 1.061(35) and 297.4(2) K in the range of 290--312 K (Fig.~\ref{fig:a-T}(b)). The change of $\beta$ occurred near 293 K with a global $T_N \approx$ 298 K. The adjusted coefficient of determination $\bar{R}^2$ was 0.990 and 0.984 for each range, respectively.
 % The values of $\beta$, $T_N$, and $\bar{R}^2$ for various fitting ranges are shown in Figs.~\ref{fig:a-T}(c) and (d).
 For the bulk-substrate sample, $\beta$ was constant at 0.324(4) with $T_N$ at 307.6 K. In the following paragraphs, we will give separate discussions to the value of $\beta$ = 0.49 in the sputtered sample, and the change of $\beta$ from 0.49 to 1.06. As a reminder to the reader, Ising, XY, and Heisenberg models rely on the sole presence of near-neighbor interactions. In the three-dimensional (3D) case, they have $\beta$ values ranging 0.32--0.38.\\
In a report based on measurements of uncompensated surface magnetization of sputtered films, values of surface $\beta$=0.5--1.0 were found \cite{fallarino_2015}. They were linked to the surface nature of the boundary magnetization in Cr\tsub{2}O\tsub{3} \cite{belashchenko_2010}, based on the assumption that $\beta$ will increase to 0.78(2) at the surface of a 3D Ising magnetic system \cite{binder_1984,landau_1990}.
% But this includes the assumption that the deeper order-parameter beneath the surface has a bulk-like behavior.
 We obtained similar values of $\beta$, but we should emphasize that $\alpha$ is a volume response and the surface contribution is negligible. According to the Ginzburg criterion \cite{pippard_1956}, the true fluctuation-dominated critical exponents are only valid in a range very close to the critical temperature. For a 3D system, that temperature range scales with the characteristic length $L$ as $L^{-6}$. Away from the critical point, the critical exponents may show a mean-field behavior. The strain is unscreened over a few hundreds of nanometers, meaning that the true exponents are present only within a few millikelvins from $T_N$. Therefore, strained films should show a mean-field behavior in the experimentally-accessible temperature resolutions \cite{scott_2008}. The presented $\beta = 0.49(1)$ is probably a mean-field exponent. Thus, we think that care should be taken in deducing the relation between $\beta$ and phase-transition classification for grown films of Cr\tsub{2}O\tsub{3} and other magnetic systems in general.\\
In a report based on measurements of exchange bias on a proximate ferromagnetic layer, a change of exchange-bias' critical exponent with temperature from 0.2--0.3 to 0.7--0.8 was found and argued to be related to the critical exponent of surface spins \cite{borisov_2011}. The exchange bias measured on a ferromagnet is a surface phenomenon, so it should infer the surface order-parameter of the AFM. However, due to the low anisotropy of Cr\tsub{2}O\tsub{3}, the exchange bias changes into an enhancement of coercivity at a temperature lower than $T_N$. The results of Ref.~\onlinecite{borisov_2011} seem to be near the blocking temperature of exchange bias, not $T_N$. On the other hand, we have an alternative explanation to the change of $\beta$ near $T_N$ from 0.49 to 1.06 in our results. We should consider the in-plane random placement of 30$^\circ$-rotated grains during sputter growth \cite{pati_2015,borisov_2016,shimomura_2016}. Two possible effects can result. The first is a randomness of $T_N$ across the film area with a normal distribution spanning 293--298 K range. The summation of the power-law contributions with different $T_N$'s can give an apparent change of $\beta$. The second possibility is lower exchange energies at grain boundaries compared to inside of grains, with the corresponding equivalent $T_N$'s of $\approx$293 K and 298 K, respectively. Then, the effective domain volume will suddenly decrease above $\approx$293 K, the thermal energy will overcome the anisotropy energy, and the fluctuations of AFM domains become dominant (\emph{i.e.}~superparamagnet-like). Within the finite-sized clusters, the order parameter is highly correlated, but the ensemble's order parameter is fluctuating in a superparamagnetic manner \cite{burke_1978}. The phase-transition should resemble a percolation process between magnetic particles in a paramagnetic matrix. The critical exponent of the order parameter of the percolation process equals exactly 1.0 in the case of a very large number of dimensions (\emph{i.e.}~mean-field-like) \cite{essam_1980}.\\
%\section{Domain reversal during field-cooling}
To check which explanation of the previous two is more probable, we measured the effect of $T^\star$, the MEFC stopping temperature, on domains population (Fig.~\ref{fig:T-star}(a)). The averaged domain state $\langle F\rangle$ of the whole sample can be estimated by normalizing the measured $\alpha_\text{peak}$ as in the following equation:
\begin{equation}
\langle F \rangle=\frac{\alpha_\text{peak}}{\alpha_\text{max}}\equiv \frac{v^+-v^-}{v^++v^-},
\end{equation}
where $v^\pm$ are the volumes of $F^\pm$ domains. MEFC was under $H_\mathrm{fr}$ = $+$10 kOe and $V_\mathrm{fr}$ = 0, $\pm$10 V. At $T^\star$ $>$ 298 K, $\langle F$(T=240 K)$\rangle$ was zero, as would be expected from zero-field cooling. At $T^\star$ $<$ 293 K, $\langle F \rangle$ was stable at $\pm$1 as expected from $\pm$MEFC. The domain state after $+$MEFC showed saturation at $T^\star$ $\leq$ 295 K, and MEFC at $V_\mathrm{fr}$ = 0 V (\emph{i.e.}~cooling with magnetic field only) showed the same behavior as $+$10 V (blue diamonds and black squares of Fig.~\ref{fig:T-star}(a)). If there is a random distribution of $T_N$, then $\langle F \rangle$-$T^\star$ should be with a width of 293--298 K. The switching after $-$MEFC showed saturation at $T^\star$ $\leq$ 293 K (red circles), with an anomalous positive pump at 296 K. If we consider that during MEFC at $H_\mathrm{fr}$ = $+$10 kOe and $V_\mathrm{fr}$ = $-$10 V there are two regions in the film: one is the cross-junction area under a MEFC($+$10 kOe, $-$10 V) preferring $F^-$ domains, and the other is the surrounding region with MEFC($+$10 kOe, 0 V) preferring $F^+$ domains. Thus, when magnetic and electric fields are being removed at $T^\star$=294--298, $F^+$ domains can propagate inside the capacitor area giving a higher population of them. $F^-$ domains are stabilized by an increase of volume-anisotropy product below 293 K. When a negative weak field $H_{<T^\star}$ of $-$20 Oe was applied below $T^\star$ to stabilize $F^-$ domains, the pump disappeared and the same behavior as $+$MEFC was obtained (magenta triangles in Fig.~\ref{fig:T-star}(a)). On the other hand, for the bulk sample, only sharp transitions at $T^\star$ = 307.5 K were observed (Fig.~\ref{fig:T-star}(b)). Thus, we consider that the more probable explanation for the change of $\beta$ is the phase-transition from an AFM state to an intermediate fluctuating state, due to a grainy structure, before becoming a PM. For both phases below $T_N$, Cr\tsub{2}O\tsub{3} remained magnetoelectric, indicating that the spin structure is preserved.\\
\section{Conclusion}
We measured the temperature dependence of magnetoelectric susceptibility in a 500-nm Cr\tsub{2}O\tsub{3} film and a bulk substrate. We found a change of critical exponent $\beta$ near $T_N$ of the Cr\tsub{2}O\tsub{3} film from 0.49 to 1.06. No such change was found in the bulk substrate.
 The magnetoelectric susceptibility is a volume measurement and cannot detect the surface component at the used thickness. So instead of an explanation based on a change of dimensionality, the films have $\beta$ value close to 0.5 due to strain resulting in a mean-field behavior. Additionally, we found that probably a phase-transition occurs within 5 K below the global N\'eel temperature from a long-range-ordered AFM to a thermally-fluctuating short-range AFM, before becoming a paramagnet above N\'eel temperature.
\begin{acknowledgments}
The authors thank Dr.~James F.~Scott for his helpful discussions. This work was partly funded by ImPACT Program of Council for Science, Technology and Innovation (Cabinet Office, Japan Government).
\end{acknowledgments}
\bibliographystyle{apsrev4-1}
\bibliography{ZotPapers,refs}

\clearpage
\onecolumngrid
\newpage

\end{document}